\documentclass[a4paper,11pt]{article}
\pdfoutput=1 

\usepackage{jheppub} 

\usepackage[T1]{fontenc} 

\title{\boldmath More on Electroweak-Skyrmion}

\author[a,b]{Ryuichiro Kitano}
\author[a]{and Masafumi Kurachi}
\affiliation[a]{Theory Center, High Energy Accelerator Research Organization (KEK), Tsukuba 305-0801, Japan}
\affiliation[b]{The Graduate University for Advanced Studies (Sokendai), Tsukuba 305-0801, Japan}

\emailAdd{Ryuichiro.Kitano@kek.jp}
\emailAdd{Masafumi.Kurachi@kek.jp}

\abstract{We discuss the properties of the topological soliton, or the Electroweak-Skyrmion, in the system of the Standard Model Higgs Lagrangian with addition of general $O(p^4)$ terms. We show that the upper bound on the mass of the Electroweak-Skyrmion is about 10~TeV, which is obtained from currently available experimental constraints on coefficients of $O(p^4)$ terms. The impact on the properties of the Electroweak-Skyrmion due to further modification of the Lagrangian is also discussed, and comments on possible mechanisms for the generation of the Electroweak-Skyrmion in the early Universe as a dark matter are given.}

\preprint{KEK-TH-1961}

\begin{document} 
\maketitle
\flushbottom

\section{Introduction}
Topological objects play important roles in various areas of physics, and the physics of the dark matter (DM) might be no exception. Actually, there have been many attempts to explain the DM as extended objects. (See, for example, Refs.~\cite{Murayama:2009nj, Ellis:2012cs}.) Recently, it was shown that even in the Standard Model (SM) with 125 GeV Higgs boson, by minimal addition of an $O(p^4)$ operator, the topological object exists as a solution of field equations~\cite{Kitano:2016ooc}. The object, Electroweak (EW)-Skyrmion, has a topological winding number in a similar manner as the Skyrmion solution in the chiral Lagrangian~\cite{Skyrme:1961vq} has it. That gives a stability to the object and makes it an attractive candidate for the DM of the Universe. Technically similar, though conceptually different, works has been published since then \cite{Matsuzaki:2016iyq, Ma:2017fjv}.

In this paper, we extend the analysis of Ref.~\cite{Kitano:2016ooc}, in which only a specific combination of $O(p^4)$ terms was considered, by allowing general $O(p^4)$ terms of the Lagrangian. Coefficients of $O(p^4)$ terms are experimentally constrained by measurements of weak gauge boson scattering amplitudes, and those constraints give an upper bound on the mass of the EW-Skyrmion. We show that the upper bound on the EW-Skyrmion mass is relaxed by about  30 - 100\% (depending on the different scattering channels) compared to the case studied in Ref.~\cite{Kitano:2016ooc}. The updated experimental analysis by the ATLAS collaboration is used to constrain the coefficients of $O(p^4)$ terms, and we show that the most stringent constraint on the mass of the EW-Skyrmion is about 10 TeV. We then study the case that the Lagrangian is further modified within the uncertainty of current experimental knowledge, and discuss the impact on the properties of the EW-Skyrmion. We also discuss the possible mechanisms for the generation of the EW-Skyrmion in the early Universe.

\section{Framework}
In this section we describe the general framework for the study of EW-Skyrmion. For the purpose of the current study, it is convenient to write the Higgs field in the matrix form as
\begin{equation}
\left( 1+\frac{h(x)}{v_{\rm EW}}\right) U(x),
\end{equation}
where
\begin{equation}
 U(x) = e^{i\, \pi^i(x)\, \sigma^i/2v_{\rm EW} }\ \ \ \ \left(\sigma^i : {\rm Pauli\ matrix}\right).
\end{equation}
Here, $\pi^i(x)$ are the Nambu-Goldstone fields associated with the electroweak symmetry breaking and $h(x)$ represents the physical Higgs boson.
The Lagrangian we study in this paper is
\begin{eqnarray}
{\cal L} &=&  {\cal G}\left(\frac{h(x)}{v_{\rm EW}} \right) \frac{v_{\rm EW}^2}{4}\, {\rm Tr}\left[  \partial_\mu U(x)^\dagger\,  \partial^\mu U(x)  \right]         \nonumber\\
           & & +\, \frac{1}{2}\partial_\mu h(x)\partial^\mu h(x) - V(h(x) )\nonumber\\
           & & +\, \alpha_4 {\rm Tr}\left[  \partial_\mu U(x)^\dagger\,  \partial_\nu U(x)  \right] {\rm Tr}\left[  \partial^\mu U(x)^\dagger\,  \partial^\nu U(x)  \right] \nonumber\\
                      & & +\, \alpha_5 \left({\rm Tr}\left[  \partial_\mu U(x)^\dagger\,  \partial^\mu U(x)  \right] \right)^2.
\label{eq:L1}
\end{eqnarray}
First and second lines in the right hand side of Eq.~(\ref{eq:L1}) become equivalent to the SM Higgs Lagrangian if we take
\begin{equation}
{\cal G}\left(\frac{h(x)}{v_{\rm EW}} \right) =  \left( 1+\frac{h(x)}{v_{\rm EW}} \right)^2
\label{eq:GSM}
\end{equation}
and
\begin{equation}
V(h(x)) = \frac{\lambda_{\rm SM}}{4} \left( (h(x)+v_{\rm EW})^2 - v_{\rm EW}^2 \right)^2,
\label{eq:VSM}
\end{equation}
where $v_{\rm EW} \simeq$ 246 GeV is the vacuum expectation value (VEV) of the Higgs field and $\lambda_{\rm SM}$ is taken to be the value which reproduces the observed Higgs mass $m_H \simeq$ 125 GeV. Terms in third and fourth lines, which are written in the notation of Ref.~\cite{Appelquist:1993ka}, represent the possible existence of anomalous quartic gauge couplings (QGCs), and coefficients ($\alpha_{4}, \alpha_{5}$) parameterize the deviation from the SM value of QGCs. Values of $\alpha_{4}$ and $ \alpha_{5}$ are constrained by measurements of scattering amplitudes of weak gauge bosons, which will be discussed in detail later. In Ref.~\cite{Kitano:2016ooc}, it was shown that the soliton solution, which we called the EW-Skyrmion, exists when we adopt the forms showin in Eqs.~(\ref{eq:GSM}), (\ref{eq:VSM}) with a specific choice of parameters, $\alpha_5=-\alpha_4$, which was motivated by the Skyrme model~\cite{Skyrme:1961vq}. The purpose of this paper is to study the properties of the EW-Skyrmion in more general Lagrangian.

EW-Skyrmion is described as a static configuration of the Higgs field with the following form:
\begin{equation}
U(x) = e^{i F(r) \sigma^i \hat{x}_i},
\label{eq:hedgehog}
\end{equation}
\begin{equation}
 h(x) = v_{\rm EW}\ \phi(r).
\end{equation}
where 
\begin{equation}
r \equiv \sqrt{x_i x_i},\ \ \ \hat{x}_i \equiv x_i/r .
\end{equation}
$F(r)$ and $\phi(r)$ can be determined by requiring that the total energy of the system is minimized. For the discussion below, it is convenient to introduce a dimensionless coordinate $\tilde{r}$ which is defined as the radial coordinate $r$ normalized by $1/v_{\rm EW}$:
\begin{equation}
\tilde{r} = r\, v_{\rm EW}.
\end{equation}
Then the energy of the static configuration can be written as
\begin{eqnarray}
E\left[ \tilde{F}(\tilde{r}), \tilde{\phi}(\tilde{r}) \right] = 
  4 \pi v_{\rm EW} \int_0^\infty d\tilde{r}\, \tilde{r}^2 &\Bigg[&
  \ \frac{1}{2}\, {\cal G}\left(\tilde{\phi}(\tilde{r})\right)
  \left( \tilde{F}'(\tilde{r})^2 + 2\, \frac{\sin^2\tilde{F}}{\tilde{r}^2} \right) \nonumber \\
  &&
  +\frac{1}{2} \tilde{\phi}'(\tilde{r})^2 + \tilde{V}\left(\tilde{\phi}(\tilde{r})\right) \nonumber \\
  &&
-4\, \alpha_4 \left( \tilde{F}'(\tilde{r})^4 + 2\, \frac{\sin^4\tilde{F}}{\tilde{r}^4} \right)\nonumber \\
  &&
-4\, \alpha_5 \left( \tilde{F}'(\tilde{r})^2 + 2\, \frac{\sin^2\tilde{F}}{\tilde{r}^2} \right)^2
 \Bigg],
\label{eq:E1}
\end{eqnarray}
where $\tilde{F}(\tilde{r})$ and $\tilde{\phi}(\tilde{r})$ are defined as
\begin{equation}
F(r) = F(\tilde{r}/v_{\rm EW}) \equiv \tilde{F}(\tilde{r}),\ \ \ \phi(r) = \phi(\tilde{r}/v_{\rm EW}) \equiv \tilde{\phi}(\tilde{r}), 
\end{equation}
and $\tilde{V}\left(\tilde{\phi}(\tilde{r})\right)$ is defined as 
\begin{equation}
V\left(v_{\rm EW} \phi(r)\right) \equiv v_{\rm EW}^4 \tilde{V}\left(\tilde{\phi}(\tilde{r})\right).
\end{equation}
The necessary condition which minimize $E[\tilde{F}(\tilde{r}), \tilde{\phi}(\tilde{r})]$ can be obtained from the Euler-Lagrange equations with respect to $\tilde{F}(\tilde{r})$ and $\tilde{\phi}(\tilde{r})$:
\begin{eqnarray}
{\cal G}\left(\tilde{\phi}(\tilde{r})\right)\left\{ \sin 2\tilde{F}(\tilde{r})  - 2\, \tilde{r}\tilde{F}'(\tilde{r}) - \tilde{r}^2\tilde{F}''(\tilde{r})\right\}
- {\cal G}' \left(\tilde{\phi}(\tilde{r})\right) \tilde{r}^2\, \tilde{\phi}'(\tilde{r})\, \tilde{F}'(\tilde{r})  
& & \nonumber\\
+16(\alpha_4+\alpha_5) \left\{ 2\, \tilde{r} \tilde{F}'(\tilde{r})^3 + 3\, \tilde{r}^2 \tilde{F}'(\tilde{r})^2 \tilde{F}''(\tilde{r})\right\}
& & \nonumber\\
-16(\alpha_4+2 \alpha_5) \frac{1}{\tilde{r}^2} \sin^2\tilde{F}(\tilde{r}) \sin2\tilde{F}(\tilde{r})
& & \nonumber\\
+16\alpha_5 \left\{ 2\, \tilde{F}''(\tilde{r}) \sin^2\tilde{F}(\tilde{r}) + \tilde{F}'(\tilde{r})^2 \sin 2\tilde{F}(\tilde{r})\right\} &=& 0. 
\label{eq:EL1}
\end{eqnarray}
\begin{eqnarray}
\frac{1}{2}\,{\cal G}' \left(\tilde{\phi}(\tilde{r})\right) \left\{  \tilde{r}^2\tilde{F}'(\tilde{r})^2 + 2 \sin^2\tilde{F}(\tilde{r}) \right\}
-2\,\tilde{r}\, \tilde{\phi}'(\tilde{r})-\tilde{r}^2 \tilde{\phi}''(\tilde{r}) 
+\, \tilde{r}^2\, \tilde{V}'\left(\tilde{\phi}(\tilde{r})\right) = 0.
\label{eq:EL2}
\end{eqnarray}
Solutions of these coupled equations are characterized by the topological winding number,
\begin{equation}
 B = - \frac{\varepsilon_{ijk}}{24 \pi^2} \int d^3x\ {\rm Tr} \Big[ V_i V_j V_k \Big],
\end{equation}
where
\begin{equation}
 V_\mu (x) \equiv \left( \partial_\mu U(x) \right)\,  U(x)^\dagger.
\end{equation}
When the system has $B=n$ ($n$: integer), $\tilde{F}(\tilde{r})$ and $\tilde{\phi}(\tilde{r})$ as solutions of Eqs.~(\ref{eq:EL1}) and (\ref{eq:EL2}) satisfy the following boundary conditions:
\begin{equation}
\tilde{F}(0) = n \pi,\ \ \ \tilde{F}(\infty) = 0.
\label{eq:BC}
\end{equation}
\begin{equation}
\tilde{\phi}'(0) = 0,\ \ \ \tilde{\phi}(\infty) = 0.
\end{equation}

In the rest of the paper, we study the system by numerically solving Eqs.~(\ref{eq:EL1}) and (\ref{eq:EL2}) with imposing the above boundary conditions. Since we are interested in the groundstate, we take $n=1$ in Eq.~(\ref{eq:BC}), and assume that the spin of the topological object is 0, therefore the value of the $E[ \tilde{F}(\tilde{r}), \tilde{\phi}(\tilde{r}) ]$ itself becomes the mass of the object when solutions are substituted back in the expression in Eq.~(\ref{eq:E1}).

\section{EW-Skyrmion in the SM with general $O(p^4)$ terms}

\begin{figure}[t]
 \begin{center}
  \includegraphics[width=80mm]{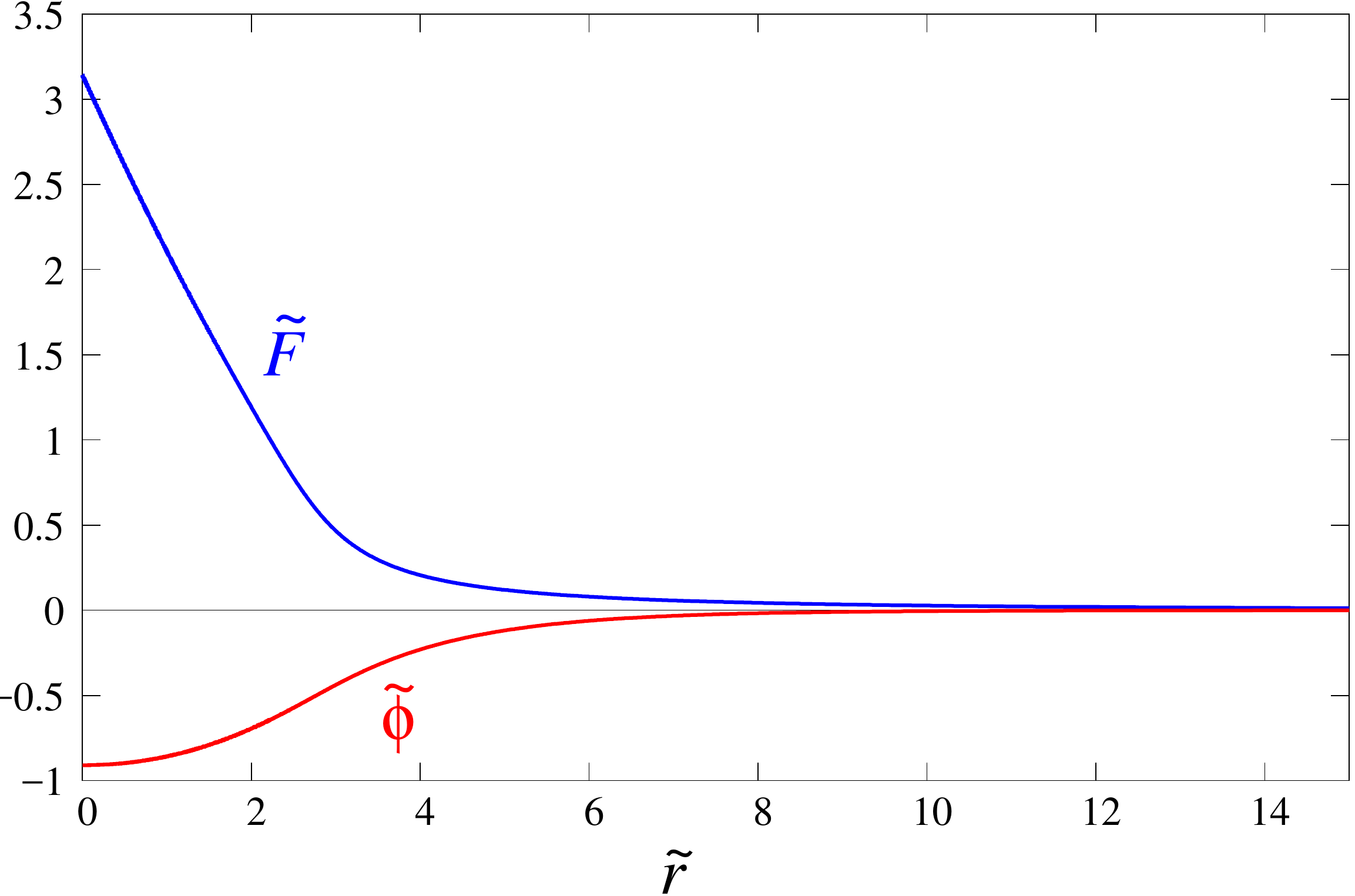}
 \end{center}
 \caption{$\tilde{F}$ (upper blue curve) and $\tilde{\phi}$ (lower red curve) as functions of $\tilde{r}$ for the case of $\alpha_4=0.1$, $\alpha_5=-0.1$. The kinetic part of the Lagrangian and the Higgs potential take the SM forms.}
 \label{fig:conf}
\end{figure}

\begin{figure}[t]
 \begin{center}
  \includegraphics[width=80mm, angle=270]{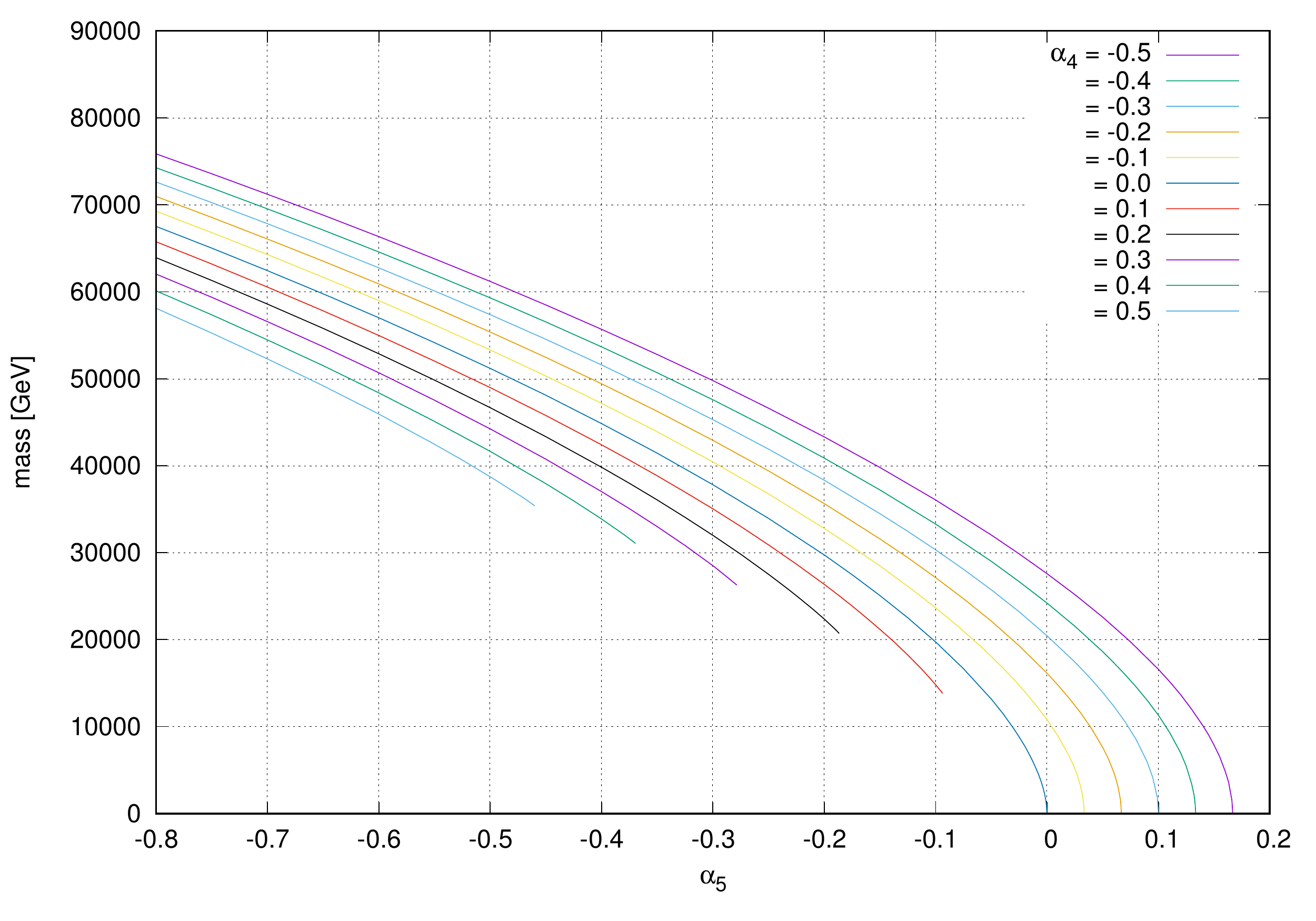} 
 \end{center}
 \caption{Mass of the EW-Skyrmion as functions of $\alpha_5$ for various choice of $\alpha_4$. Kinetic part of the Lagrangian and the Higgs potential are taken to be the SM forms (Eqs.~(\ref{eq:GSM}), (\ref{eq:VSM})).}
 \label{fig:mass-SM}
\end{figure}

In this section, we assume that the kinetic part of the Lagrangian and the Higgs potential take the SM forms, namely we adopt the expression shown in Eqs.~(\ref{eq:GSM}), (\ref{eq:VSM}), and study the effect of the $O(p^4)$ terms by varying parameters $\alpha_4$ and $\alpha_5$. In Fig.~\ref{fig:conf}, we show the example of numerical solutions for $\tilde{F}(\tilde{r})$ (upper blue curve) and $\tilde{\phi}(\tilde{r})$ (lower red curve) in the case of $\alpha_4=0.1$, $\alpha_5=-0.1$. %
As mentioned at the end of the previous section, the mass of the EW-Skyrmion is obtained by substituting these solutions back into the expression of $E[ \tilde{F}(\tilde{r}), \tilde{\phi}(\tilde{r}) ]$ in Eq.~(\ref{eq:E1}). We calculated the mass of the EW-Skyrmion in the range of $-0.5 \leq \alpha_4 \leq 0.5$ and $\alpha_5 \geq -0.8$, and plot the results in Fig.~\ref{fig:mass-SM}. In the figure, the mass of the EW-Skyrmion is plotted as a function of $\alpha_5$ for various choice of $\alpha_4$.
Note that for $\alpha_4 \leq 0$, the mass of the EW-Skyrmion monotonically decreases as $\alpha_5$ becomes large, and goes to $0$ in the limit of approaching  a certain critical value of $\alpha_5$. In the region where $\alpha_5$ is larger than this critical value, there is no solution which satisfies the required equations and boundary conditions. In the case of $\alpha_4 > 0$, the decreasing tendency as a function of $\alpha_5$ is the same as in the case of $\alpha_4 \leq 0$, however, the solution disappears before it reaches to the massless limit. This is because the necessary condition for the existence of the minimum of Energy functional $E[\tilde{F}(\tilde{r}), \tilde{\phi}(\tilde{r})]$ with respect to the variation of $\tilde{F}(\tilde{r})$ and $\tilde{\phi}(\tilde{r})$ (Legendre's condition) is not satisfied in larger $\alpha_5$ region.

\begin{figure}[t]
 \begin{center}
  \includegraphics[width=130mm]{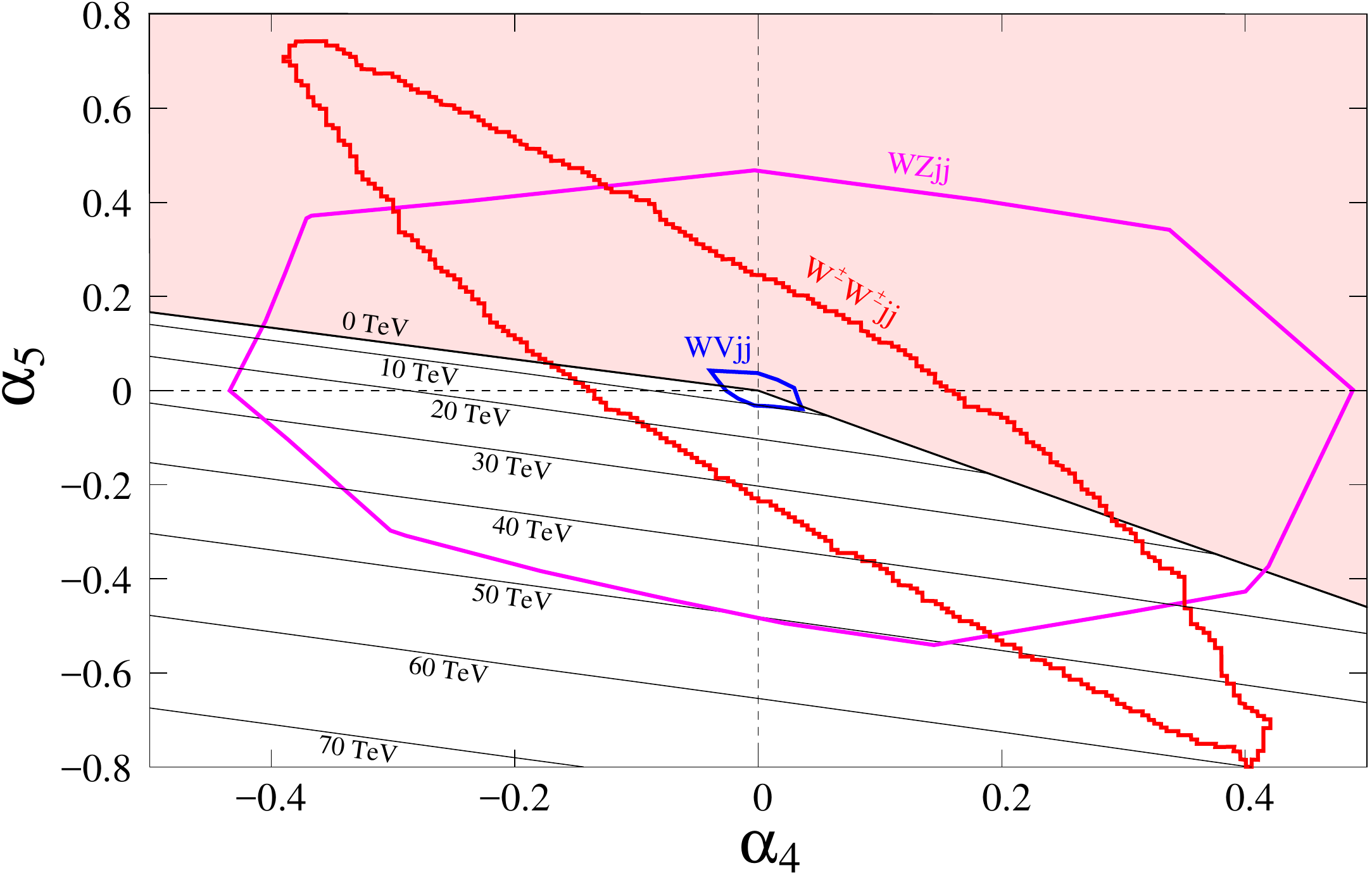}
 \end{center}
 \caption{Mass of the EW-Skyrmion on the $\alpha_4$-$\alpha_5$ plane. Contours for the mass of 10 TeV, 20 TeV, $\cdots$ 70 TeV are shown as black solid lines. In the shaded regions, there is no solution which satisfies required equations and boundary conditions. 95\% CL allowed region (inside of contours) from several experiments (Refs.\cite{Aad:2014zda,Aad:2016ett,Aaboud:2016uuk}), indicated by $W^\pm W^\pm jj$, $WZjj$ , and $WVjj$, are also shown. (See the main text for detailed explanations.)}
 \label{fig:constraint}
\end{figure}

In Fig.~\ref{fig:constraint} the mass derived above is shown on the $\alpha_4$-$\alpha_5$ plane. 
In the shaded regions, there is no solution which satisfies required equations and boundary conditions. In the un-shaded region, contours for the mass of 10 TeV, 20 TeV, $\cdots$ 70 TeV are illustrated  as black solid lines. There are two distinct boundaries which separates the shaded and un-shaded regions in $\alpha_4 \leq 0$ and $\alpha_4 > 0$, respectively, which match the explanation in the previous paragraph: in the region of $\alpha_4 \leq 0$, approaching the boundary from below, the mass of the EW-Skyrmion goes to 0, meanwhile, in the region of $\alpha_4 >0$, the solution disappears at the boundary with finite value of the mass.

As we mentioned in the previous section, $\alpha_{4}$ and $\alpha_{5}$ parameterize the deviation from the SM value of QGCs, and magnitude of them are experimentally constrained by measurements of weak gauge boson scattering amplitudes. In Fig.~\ref{fig:constraint}, we overlay the currently available constraints from ATLAS analyses (Refs.~\cite{Aad:2014zda,Aad:2016ett,Aaboud:2016uuk}) of LHC experiment. The contours indicated by $W^\pm W^\pm jj$, $WZjj$, and $WVjj$ in the figure are 95\% CL limits (inside of them are allowed region) from various channels of weak gauge boson scattering measurements: those indicated by $W^\pm W^\pm jj$ \cite{Aad:2014zda} and $WZjj$ \cite{Aad:2016ett} used leptonic final states of same-sign $WW$ scattering and $WZ$ scattering, respectively, while the one indicated by $WVjj$ \cite{Aaboud:2016uuk} used semi-leptonic final states, where $W$ decays leptonically and $V$ ($=W$ or $Z$) decays hadronically. (``$jj$'' in the name of all channels means forward jets associated with weak gauge boson fusion process.) From this figure, the upper bound on the mass of the EW-Skyrmion, $M$, constrained from each analysis can be derived as follows:
\begin{eqnarray}
 W^\pm W^\pm jj \ &:&\ \  M \lesssim 60.1\, {\rm TeV}\ \ \ (\, \alpha_4 \simeq 0.4,\, \alpha_5 \simeq -0.8 \,), \label{eq:Mub1}\\
 WZ jj \ \ \  &:&\ \  M \lesssim 50.5\, {\rm TeV}\ \ \ (\, \alpha_4 \simeq 0.14,\, \alpha_5 \simeq -0.54 \,), \label{eq:Mub2}\\
 WV jj\ \ \ &:&\ \  M \lesssim 10.3\, {\rm TeV}\ \ \ (\, \alpha_4 \simeq -0.003,\, \alpha_5 \simeq -0.032 \,),\label{eq:Mub3} 
\end{eqnarray}
where the values of $(\alpha_4, \alpha_5)$ which give the largest $M$ in the allowed region in each channel are indicated in parentheses. 

It is interesting to compare above results to the case where values of $\alpha_4$ and $\alpha_5$ are constrained by the relation $\alpha_5=-\alpha_4$. As briefly mentioned earlier, this specific choice of parameters give a Lagrngian term which has the same form as that introduced by the original Skyrme model~\cite{Skyrme:1961vq}. This can be easily seen by rewriting the $O(p^4)$ terms in the Lagrangian as follows:
\begin{eqnarray}
{\cal L}_{{\cal O}(p^4)} &=&  \alpha_4 {\rm Tr}\left[  \partial_\mu U(x)^\dagger\,  \partial_\nu U(x)  \right] {\rm Tr}\left[  \partial^\mu U(x)^\dagger\,  \partial^\nu U(x)  \right] +\, \alpha_5 \left({\rm Tr}\left[  \partial_\mu U(x)^\dagger\,  \partial^\mu U(x)  \right] \right)^2, \nonumber\\
&=&
\alpha_4\, {\rm Tr}\left[ V_\mu V_\nu \right] {\rm Tr}\left[ V^\mu V^\nu \right]
                  \ +\        \alpha_5\, {\rm Tr}\left[ V_\mu V^\mu \right] {\rm Tr}\left[ V_\nu V^\nu \right], \nonumber \\
&=&-\frac{1}{2} \, \alpha_5 \,{\rm Tr}\Big[ \left[V_\mu (x) , V_\nu(x)\right] \left[V^\mu (x) , V^\nu(x)\right]  \Big] \nonumber \\
&& \ \ \ +\,\frac{1}{2} \, (\alpha_4 + \alpha_5) \,{\rm Tr}\Big[ \left\{V_\mu (x) , V_\nu(x)\right\} \left\{V^\mu (x) , V^\nu(x)\right\}  \Big],
\end{eqnarray}
where $[A, B] \equiv AB - BA$ and $\{A, B\} \equiv AB + BA$, respectively. The first term in the last expression above is nothing but the Skyrme term, and the secont term drops when one takes $\alpha_5=-\alpha_4$.  The upper bounds on the mass when we restrict the parameter in this way are summarized as follows:
\begin{eqnarray}
 W^\pm W^\pm jj \ &:&\ \  M \lesssim 29.0\, {\rm TeV}\ \ \ (\, \alpha_4 = -\alpha_5 \simeq 0.31 \,), \label{eq:Mub4}\\
 WZ jj \ \ \  &:&\ \  M \lesssim 34.4\, {\rm TeV}\ \ \ (\, \alpha_4 = -\alpha_5 \simeq 0.41 \,), \label{eq:Mub5}\\
 WV jj\ \ \ &:&\ \  M \lesssim 8.0\, {\rm TeV}\ \ \ \ \, (\, \alpha_4 = -\alpha_5 \simeq 0.035 \,), \label{eq:Mub6}
\end{eqnarray}
again, values of $(\alpha_4, \alpha_5)$ which give the largest $M$ in the allowed region, namely the cross section of the line $\alpha_5=-\alpha_4$ and 95\% CL contour of each experimental analysis, are indicated in parentheses. From these, it is concluded that the upper bound on the mass of EW-Skyrmion is relaxed by about 100\%, 50\%, 30\% for the case of $W^\pm W^\pm jj$, $WZjj$, and $WVjj$ channels, respectively, when one allows general choice of $\alpha_4$ and $\alpha_5$ compared to the case where those are restricted to $\alpha_4=-\alpha_5$ as in the form of the Skyrme model.


Before move on to the next section, let us briefly discuss the comparison of the result here to the case that no scalar particle is included in the analysis, which is equivalent to the limit that Higgs mass is taken to be infinity with fixing the value of $v_{\rm EW}$. (Such analysis, namely the study of the Skyrmion solution in the context of non-linear EW chiral Lagrangian has been done in Refs.~\cite{Ellis:2012cs}.)
\begin{figure}[t]
 \begin{center}
  \includegraphics[width=60mm, angle=270]{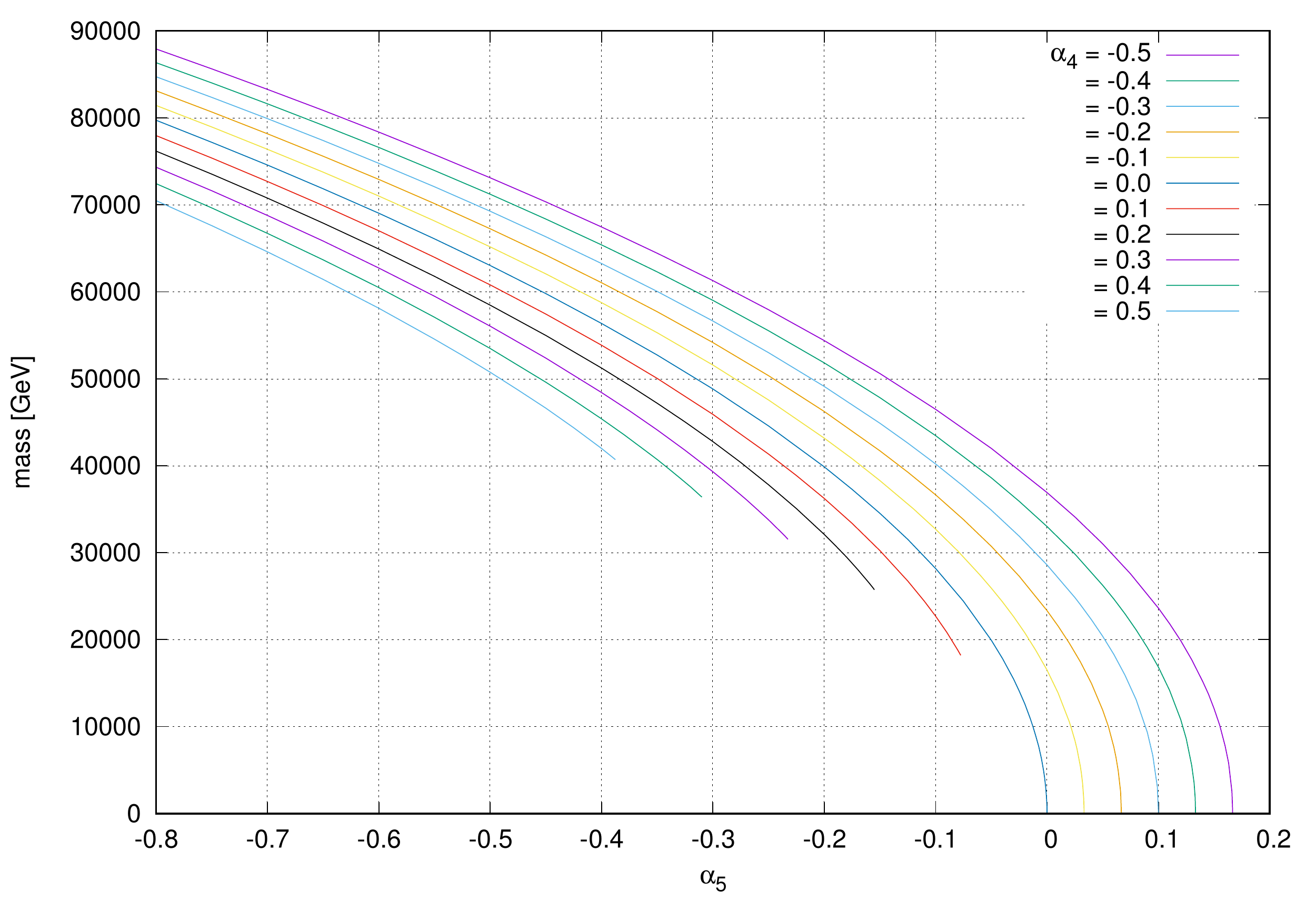}
 \end{center}
 \caption{Mass of the EW-Skyrmion for the case that no scalar field is included. The mass is plotted as functions of $\alpha_5$ for various choice of $\alpha_4$.}
 \label{fig:mass-NoSC}
\end{figure}
The mass of the EW-Skyrmion in this case is plotted as functions of $\alpha_5$ for various choice of $\alpha_4$ in Fig.~\ref{fig:mass-NoSC}. In Ref.~\cite{Kitano:2016ooc}, for the specific case of $\alpha_5 = - \alpha_4$, it was shown that inclusion of the scalar field had an effect that the mass of the EW-Skyrmion is reduced compared to the no-scalar case. (This is also pointed out in Ref.~\cite{He:2015eua} in the context of hadron physics.)  By comparing the result in Fig.~\ref{fig:mass-NoSC} to that in Fig.~\ref{fig:mass-SM}, one can see that the effect of reducing the EW-Skyrmion mass by the inclusion of the scalar field is general tendency for arbitrary choice of $\alpha_4$ and $\alpha_5$ at least in the region investigated here.

\section{Further generalizations}
In this section, we study the properties of the EW-Skyrmion by generalizing the form of the Lagrangian further. In the previous section, we assumed that the function ${\cal G}\left(\frac{h(x)}{v_{\rm EW}} \right)$ and the Higgs potential $V(h(x))$ take the SM forms, namely Eqs.~(\ref{eq:GSM}), (\ref{eq:VSM}), however, these forms are not established experimentally yet, and there is a possibility that deviation from the SM forms will be found at future experiments. Therefore, we study the effect of the modifications on the properties of the EW-Skyrmion. 

For the function ${\cal G}\left(\frac{h(x)}{v_{\rm EW}} \right)$, we consider the following form as an extension:
\begin{equation}
{\cal G}\left(\frac{h(x)}{v_{\rm EW}} \right) =  1+ 2\,a\left(\frac{h(x)}{v_{\rm EW}}\right) + b \left( \frac{h(x)}{v_{\rm EW}} \right)^2.
\label{eq:Ggen}
\end{equation}
As for the Higgs potential, we modify it by adding dimension-6 operator:
\begin{equation}
V(h(x)) = \frac{\lambda_{\rm SM}}{4} \left( (h(x)+v_{\rm EW})^2 - v_{\rm EW}^2 \right)^2
+ c\,\frac{\lambda_{\rm SM}}{8 v_{\rm EW}^2} \left( (h(x)+v_{\rm EW})^2 - v_{\rm EW}^2 \right)^3.
\label{eq:Vgen}
\end{equation}
By the addition of the last term, the triple Higgs coupling is modified as $v_{\rm EW}\lambda_{\rm SM} \rightarrow v_{\rm EW}\lambda_{\rm SM}(1+c)$, while keeping the Higgs mass and the Higgs VEV unchanged. The choice of the parameters $(a, b, c) = (1, 1, 0)$ (with $(\alpha_4, \alpha_5)=(0,0)$) reproduces the SM Higgs Lagrangian.

\begin{figure}[t]
 \begin{center}
  \includegraphics[width=52mm, angle=270]{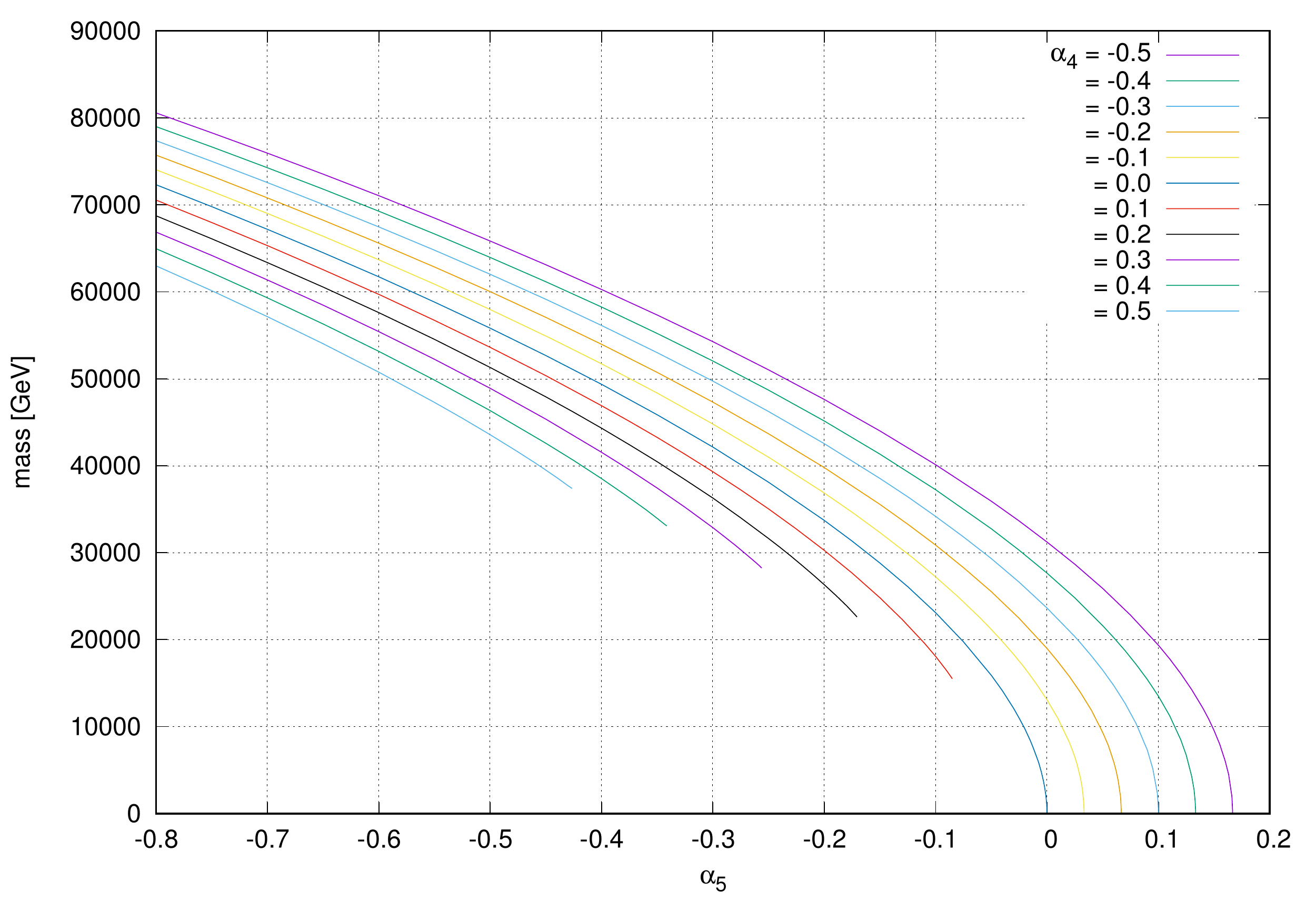}\ \ 
  \includegraphics[width=52mm, angle=270]{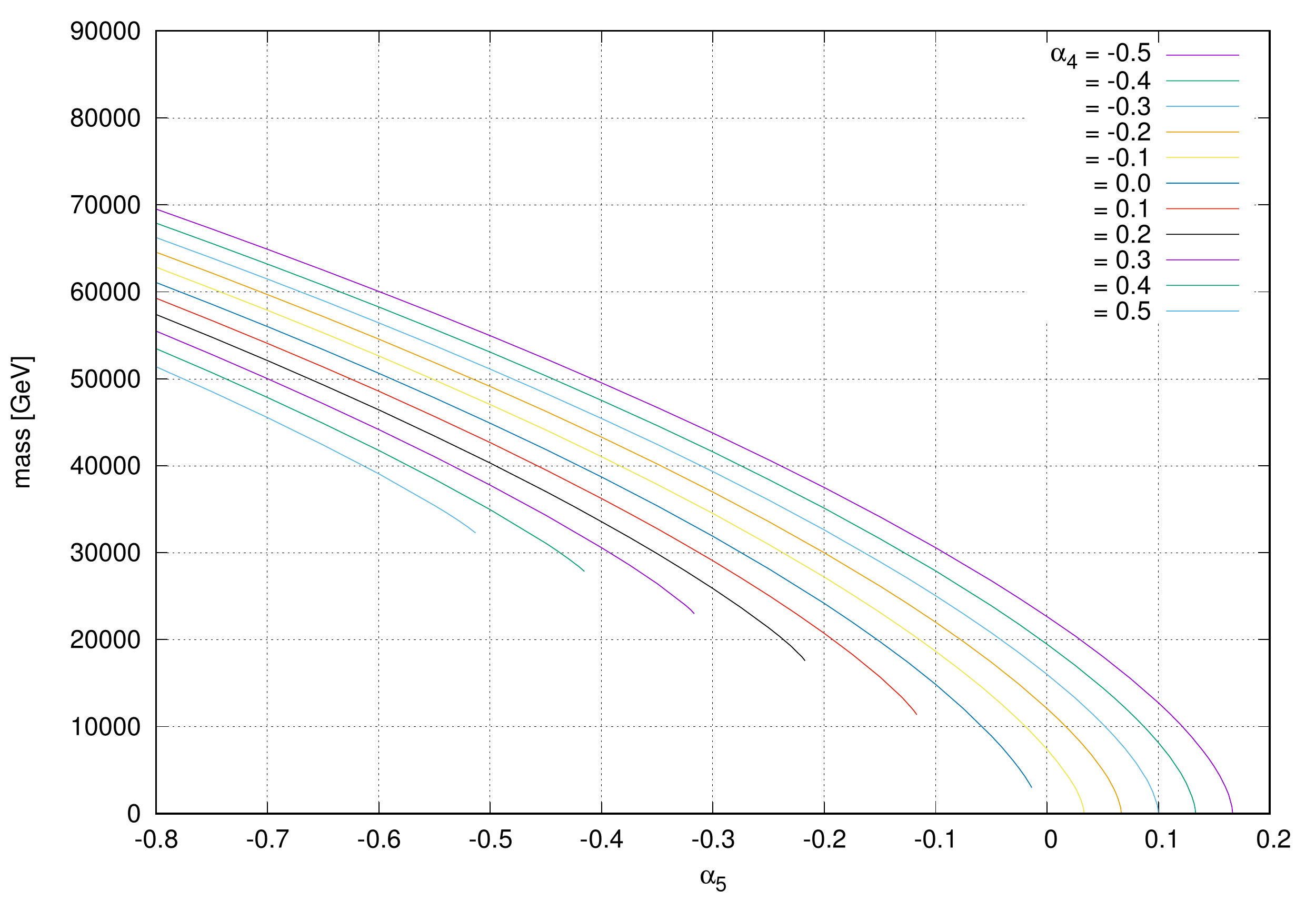}
 \end{center}
 \caption{Mass of the EW-Skyrmion as functions of $\alpha_5$ for various choice of $\alpha_4$ for the cases of $(a, b, c) = (0.8, 1, 0)$ (left panel) and $(a, b, c) = (1.2, 1, 0)$ (right panel).}
 \label{fig:mod-a}
\end{figure}

We first study the effect of the change of the parameter $a$. Here, we consider two cases, $a=0.8$ and $1.2$, namely the effect of the $\pm$ 20\% deviations from the SM value. The reason for this choice beeing that from the measurement of the Higgs coupling to weak gauge bosons \cite{Khachatryan:2016vau}, it is unlikely that the deviation from the SM value is much larger than this amount. In Fig.~\ref{fig:mod-a}, the mass of the EW-Skyrmion is plotted in a similar manner to Fig.~\ref{fig:mass-SM} for the case of $(a, b, c) = (0.8, 1, 0)$ (left panel) and $(a, b, c) = (1.2, 1, 0)$ (right panel).
From these plots, by comparing to Fig.~\ref{fig:mass-SM}, one can see that smaller (larger) value of $a$ has an effect to enhance (reduce) the mass of the EW-Skyrmion. It is also interesting to note that, in the positive $\alpha_4$ region, the critical value of the $\alpha_5$ above which the solution disappears become smaller (larger) when $a$ is taken to be larger (smaller) than $1$. Due to this, in the case of $a=1.2$ (right panel), the critical value becomes smaller than $-\alpha_4$, therefore, in this case, there is no solution if one takes the Skyrme-model like parameter choice $\alpha_5 = -\alpha_4$.

\begin{figure}[t]
 \begin{center}
  \includegraphics[width=52mm, angle=270]{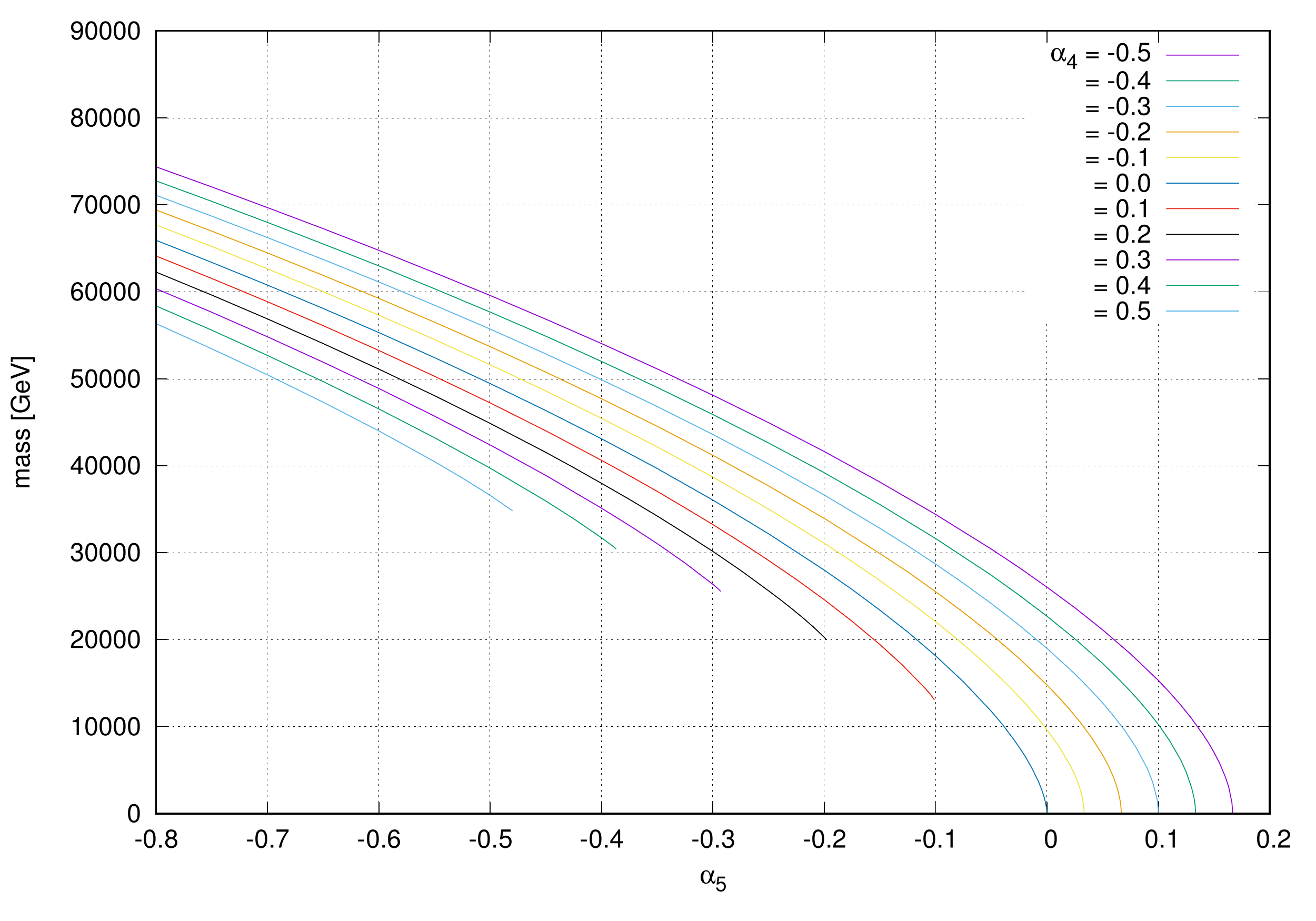}\ \ 
  \includegraphics[width=52mm, angle=270]{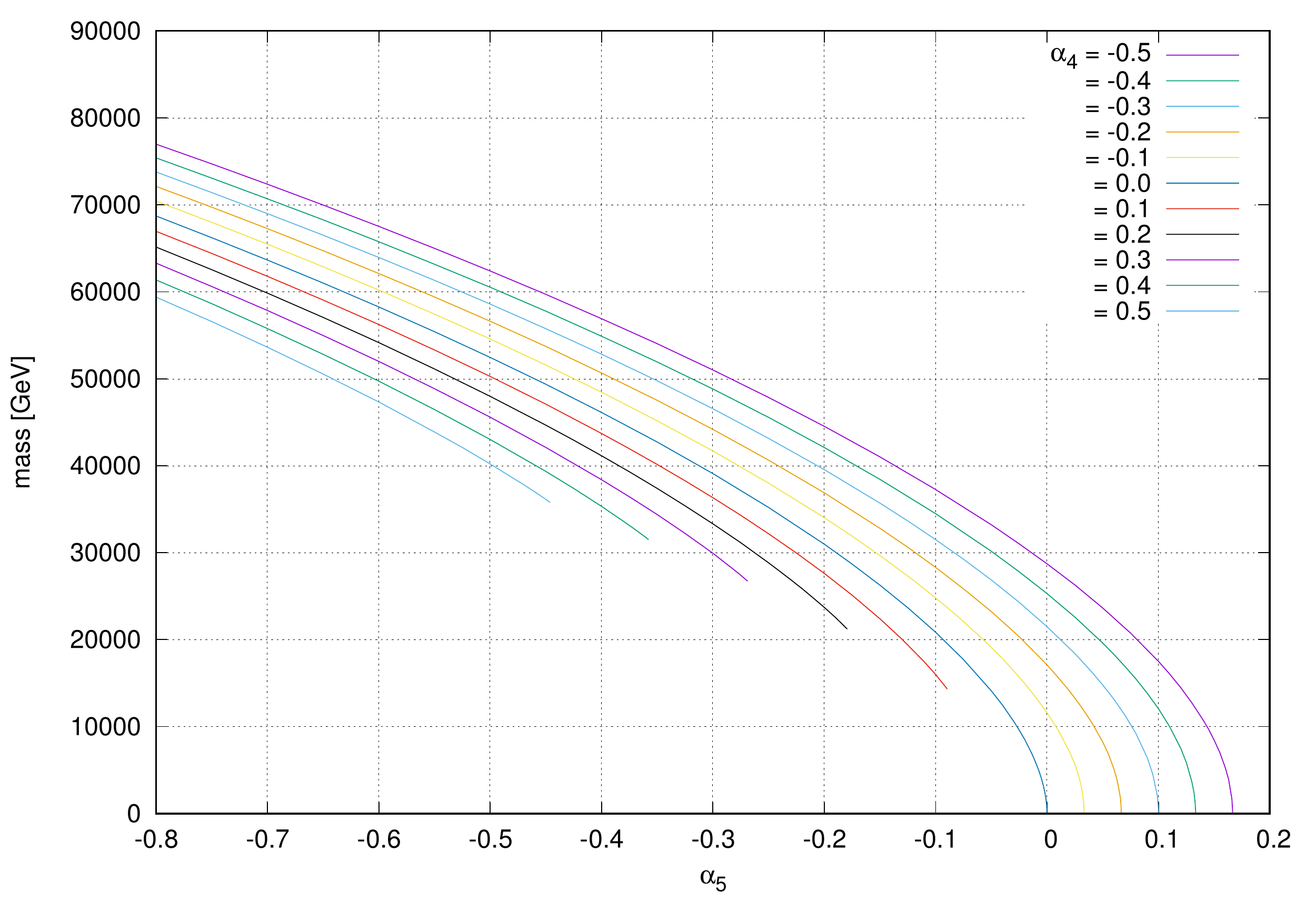}
 \end{center}
 \caption{Mass of the EW-Skyrmion as functions of $\alpha_5$ for various choice of $\alpha_4$ for the cases of $(a, b, c) = (1, 0.8, 0)$ (left panel) and $(a, b, c) = (1, 1.2, 0)$ (right panel).}
 \label{fig:mod-b}
\end{figure}

Next, we study the effect of the change of parameters $b$ and $c$. Unlike for the case of $a$, there are no stringent experimental constraint for these parameters yet. Here, to see the impact of the change of these parameters, we take 20\% deviation from the SM value as a benchmark. In Fig.~\ref{fig:mod-b}, the mass of the EW-Skyrmion is plotted for the cases of $(a, b, c) = (1, 0.8, 0)$ (left panel) and $(a, b, c) = (1, 1.2, 0)$ (right panel).
By comparing these plots to Fig.~\ref{fig:mass-SM}, one can see that smaller (larger) value of $b$ has an effect to reduce (enhance) the mass of the EW-Skyrmion.
We have done the similar analysis for the case of $(a, b, c) = (1, 1, 0.2)$, namely the 20\% change of higgs triple coupling, and it turned out that the effect is tiny and the difference from the case of $(a, b, c) = (1, 0.8, 0)$ is almost negligible.

In a similar way used in the previous section, we have derived upper bound on the mass of the EW-Skyrmion for the case of parameter choices studied in this section. The results are summarized in Table~\ref{tab:table}.
\begin{table}[htb]
\begin{center}
  \begin{tabular}{|c|c|c|c|c|c|c|} \hline
    parameter  & \multicolumn{2}{c|}{$W^\pm W^\pm$ (all leptonic)} &\multicolumn{2}{c|}{$W^\pm Z$ (all leptonic)} &\multicolumn{2}{c|}{$W^\pm V$ (semi-leptonic)}  \\ \cline{2-7}
(a, b, c) &  Skyrme & General & Skyrme & General & Skyrme & General \\ \hline \hline
(1.0,\, 1.0,\, 0.0) & 29.0 & 60.1 & 34.4 & 50.5 & 8.0 &10.3 \\ \hline
(0.8,\, 1.0,\, 0.0) & 33.5 & 65.0 & 39.0& 55.2 & 10.3 & 12.7 \\ \hline
(1.2,\, 1.0,\, 0.0) & -- & 53.5 & -- & 44.2 & -- & 6.5 \\ \hline
(1.0,\, 0.8,\, 0.0) & 26.9 & 58.4 & 32.2 & 48.8 & -- & 9.1 \\ \hline
(1.0,\, 1.2,\, 0.0) & 30.5 & 61.4 & 35.8 & 51.8 & 8.9 & 11.2 \\ \hline
(1.0,\, 1.0,\, 0.2) & 28.9 & 59.8 & 34.1 &  50.2 & 8.0 & 10.3 \\ \hline \hline
 No-scalar & 40.0 & 72.4 & 45.9 & 62.4 & 13.4 & 16.2 \\ \hline
  \end{tabular}
  \caption{Upper bound on the mass of the EW-Skyrmion in the unit of TeV. Bounds for the case of the Lagrangian without scalar field are also shown for comparison.}
\label{tab:table}  
\end{center}
\end{table}
As we have done in the previous section, in addition to deriving upper bounds when we allow general values of $\alpha_4$ and $\alpha_5$, we also derived those when we restrict the parameters to be Skyrme-model like (namely, $\alpha_5=-\alpha_4$). The values of $(\alpha_4, \alpha_5)$ which give the largest $M$ in the allowed region are the same as those in the case of $(a,b,c)=(1,1,0)$, which were shown in Eqs.~(\ref{eq:Mub1})-(\ref{eq:Mub3}) and (\ref{eq:Mub4})-(\ref{eq:Mub6}). ``--'' in the table means there is no Skyrmion solution in the allowed region. In addition to the bounds derived in this section, those derived in the previous section, namely for the case of $(a,b,c)=(1,1,0)$ are shown on the top row of the table. Also, for comparison,  similar mass bounds for the case that no scalar particle is included in the analysis are shown as well.

\section{Cosmology}

We here comment on the possible mechanisms for the generation of the Skyrmion dark matter in the early Universe. Although the EW-Skyrmion considered in this paper is classically stable due to its non-trivial topological charge, the instanton in the $SU(2)_L$ gauge theory may unwind the Skyrmion. In the early Universe, when the temperature is above the electroweak phase transition, the spharelon process can efficiently convert the Skyrmion into the baryons or leptons in the SM.

The situation is, however, quite model dependent. By analogy of the Skyrmion in QCD as the nucleon, the EW-Skyrmion would be identified with the techni-baryon like object in the full dynamics of the electroweak symmetry breaking. In this case, depending on the structures of the full theory, the techni-baryon can carry its own baryon number which is not anomalous to the $SU(2)_L$ gauge theory. This situation is possible if there are multiple techni-baryons so that the anomaly is canceled.
\footnote{We thank Masahiro Takimoto for discussion on this point.}

There are, therefore, two kinds of scenarios one can consdier; the case with the stable EW-Skyrmion and the other with the EW-Skyrmion unstable under the electroweak spharelon process.
The case with stable EW-Skyrmion is discussed in Ref.~\cite{Kitano:2016ooc}. Because of the large annihilation cross section, it is difficult to explain the abundance by the thermal relic. An interesting possibility is to assume that the Skyrmion asymmetry generated in the early Universe has remained today as the dark matter.

If the EW-Skyrmion is unstable under the spharelon process, even the asymmetry cannot survive if the reheating temperature of the Universe exceeds the electroweak scale. In order to explain the abundance today, the reheating temperature should be lower than the electroweak scale.
In this case, the non-thermal production of the Skyrmion is possible if the inflaton is directly coupled to the EW-Skyrmion. The number density to the entropy ratio is estimated by $n/s \sim (T_R / m_\phi) Br$, where $T_R$, $m_\phi$, $Br$ are the reheating temperature, the inflaton mass, and the branching ratio into the EW-Skyrmion, respectively. If the generated abundance is high enough, the annihilation process reduces the abundance to $n/s \sim (\langle \sigma v \rangle T_R M_{\rm
Pl})^{-1}$~\cite{Moroi:1999zb}. A large enough amount of the EW-Skyrmion can be generated even if the annihilation process is very strong.

Another possibility is to assume the production of the EW-Skyrmion during the reheating process. The decay products of the inflaton can hit particles in the thermal plasma, and heavy particles can be produced by this process even if the reheating temperature is very low~\cite{Harigaya:2014waa}. For the Skyrmion mass around 10~TeV, the reheating temperature of the order of 100~MeV can explain the abundance.

\section{Summary}

The discovery of the Higgs particle was definitely a major step forward for our understanding of the nature of the elementary particle physics. The next step is to study the properties of the Higgs itself. In this paper, we pursued the possibility that the Higgs field plays an important role, not only in the form of a particle, but also in the form of topological objects. General $O(p^4)$ terms were added to the SM Higgs Lagrangian, and it was shown that topologically non-trivial configurations exist as solutions of field equations. The upper bound on the mass of such object, the EW-Skyrmion, was estimated to be about 10~TeV, and the impact on the properties of the EW-Skyrmion due to further modifications of the Lagrangian was discussed. The EW-Skyrmion can be identified as the DM, and possible mechanisms for the generation of the EW-Skyrmion in the early Universe was discussed. The experimental constraints on Lagrangian parameters discussed in this paper are expected to become significantly more precise near future, and there is a possibility that deviation from the SM is discovered, in which case the analysis done in this paper will provide a prediction of the mass of the EW-Skyrmion or disprove the existence of such topological objects depending on the values of parameters.

\acknowledgments
We thank Keisuke Harigaya for discussions. This work was supported by JSPS KAKENHI Grant-in-Aid for Scientific Research (B) (No. 15H03669 [RK]) and MEXT KAKENHI Grant-in-Aid for Scientific Research on Innovative Areas (No. 25105011 [RK,MK]).

\end{document}